\documentclass[noshowpacs,prl,aps,superscriptaddress,floatfix,twocolumn]{revtex4-2} 
\usepackage[utf8]{inputenc}
\usepackage{lineno,mathtools,url}
\usepackage{amsmath}
\usepackage{amssymb,mathrsfs}
\usepackage{physics}
\usepackage{gensymb}
\usepackage{graphicx}
\usepackage{hyperref}
\hypersetup{
     colorlinks=true,
     linkcolor=blue,
     filecolor=blue,
     citecolor = orange,      
   }
\usepackage[dvipsnames]{xcolor}
\usepackage{dsfont}
\usepackage{soul}
\usepackage{mathbbol} 
\usepackage[normalem]{ulem}
\allowdisplaybreaks
\begin{document}

\title{An optically accelerated extreme learning machine using hot atomic vapors}

\author{P. Azam}
\author{R. Kaiser}
\affiliation{Université Côte d'Azur, CNRS, Institut de Physique de Nice, F-06200 Nice, France}
\begin{abstract}
Machine learning is becoming a widely used technique with a impressive growth due to the diversity of problem of societal interest where it can offer practical solutions. This increase of applications and required resources start to become limited by present day hardware technologies. Indeed, novel machine learning subjects such as large language models or high resolution image recognition raise the question of large computing time and energy cost of the required computation. In this context, optical platforms have been designed for several years with the goal of developing more efficient hardware for machine learning. Among different explored platforms, optical free-space propagation offers various advantages: parallelism, low energy cost and computational speed.
Here, we present a new design combining the strong and tunable nonlinear properties of a light beam propagating through a hot atomic vapor with an Extreme Learning Machine model.
We numerically and experimentally demonstrate the enhancement of the training using such free-space nonlinear propagation on a MNIST image classification task. We point out different experimental hyperparameters that can be further optimized to improve the accuracy of the platform.
\end{abstract}

\maketitle

\section{Introduction}
\noindent Machine learning is commonly used at a global scale in a extremely wide range of applications, offering powerful tools in many domains. Due to their adaptability and efficiency, artificial neural networks (ANNs) models have been widely implemented in most of machine learning applications. The performances provided by such models can be explained by the large number of training parameters that can be implemented, reaching several billions for the biggest models ($175$ billions for GPT-3 \cite{Floridi20}). These models, even though powerful in terms of accuracy, require important computational resources to perform training which yields a heavy cost both in time and energy. The increase of complexity in machine learning tasks due to the democratisation and spread of such technologies leads to new global issues such as excessive energy consumption of these models with impact on global warming \cite{McCallum_20}. In addition, specific applications require high computation speed and have intrinsic energy consumption constraints \cite{Hanzlik21, Mie22} .\\
Alternative methods are therefore developed to tackle these issues: one of them aims to use physical systems as fast and energy efficient platforms to process the data transformation required for machine learning. In this context photonic platforms appear to be a good candidate offering a fast, energy efficient and scalable solution \cite{wetzstein20} in the time domain \cite{lupo21, yildirim22, Fischer23} as well as in the spatial domain \cite{bueno18, lin18, Zuo19, pierangeli21, bu22}.
Methods going beyond those used by usual neural networks models such as transformers models \cite{kamath22} have been developed in order to parallelize information process and perform training on larger data sets and are widely use in natural language processing \cite{nadkarni11} and computer vision \cite{jahne00}. This approach is well adapted for optical platforms in particular in the spatial domain where large data can be processed in parallel using simple devices such as digital micromirror devices (DMD) or spatial light modulators (SLM).\\
Several optical platforms are already used for machine learning tasks, with present investigations focusing on the  identification of specific potential to be competitive for precise user cases \cite{mcmahon23}.
Among the various possibilities, a powerful method to adapt optical platforms for such applications is based on reservoir computing \cite{Schrauwen07, Vander17}. Unlike ANNs, where data is transformed by a trained nonlinear system and then read, the reservoir computing model consists in the transformation of data by a fixed and unknown nonlinear system while the training occurs at the readout. This conceptual difference with ANNs, despite being applicable to optical platforms, is also highly efficient in terms of cost: the number of required training parameters being several orders of magnitude smaller. Optical based reservoir computing models have shown promising performances using silicon chips \cite{vandoorne14,bandyopadhyay22}, laser networks \cite{Rohm20}, disordered media \cite{xia23}, Kerr nonlinear media \cite{Pauwels19, Marcucci20, xu20, Psaltis21, Moss22, boikov23} or delayed-based systems \cite{duport12, paquot12, chembo20}. 
Having a similar structure, Extreme Learning Machines (ELMs) \cite{HUANG06, Wang22} are composed of a single hidden feed-forward layer performing a nonlinear transformation of the data with a training done at the readout. Unlike reservoir computing however, the reservoir used in ELMs is not designed to process dynamical memory. ELM models are based on support vector machines \cite{noble06} and kernel methods \cite{hofmann08} and can be implemented on   various optical platforms. Among these different platforms, hot atomic vapors are a good candidate offering strong nonlinearities and are suitable for free-space propagation. We note that this nonlinear medium is well known in academic research and used in multiple fields of research \cite{martin18, glorieux23} including machine learning \cite{ryou21, silva24}. An important advantage using hot atomic vapors is the possibility to tune the strength of the nonlinear transformation, giving access to a critical parameter to optimize the learning for a given problem.\\

In this paper we present the use of such hot atomic vapor as a tunable nonlinear optical medium for machine

\begin{figure*}[!t]
\centering
\includegraphics[width=\linewidth]{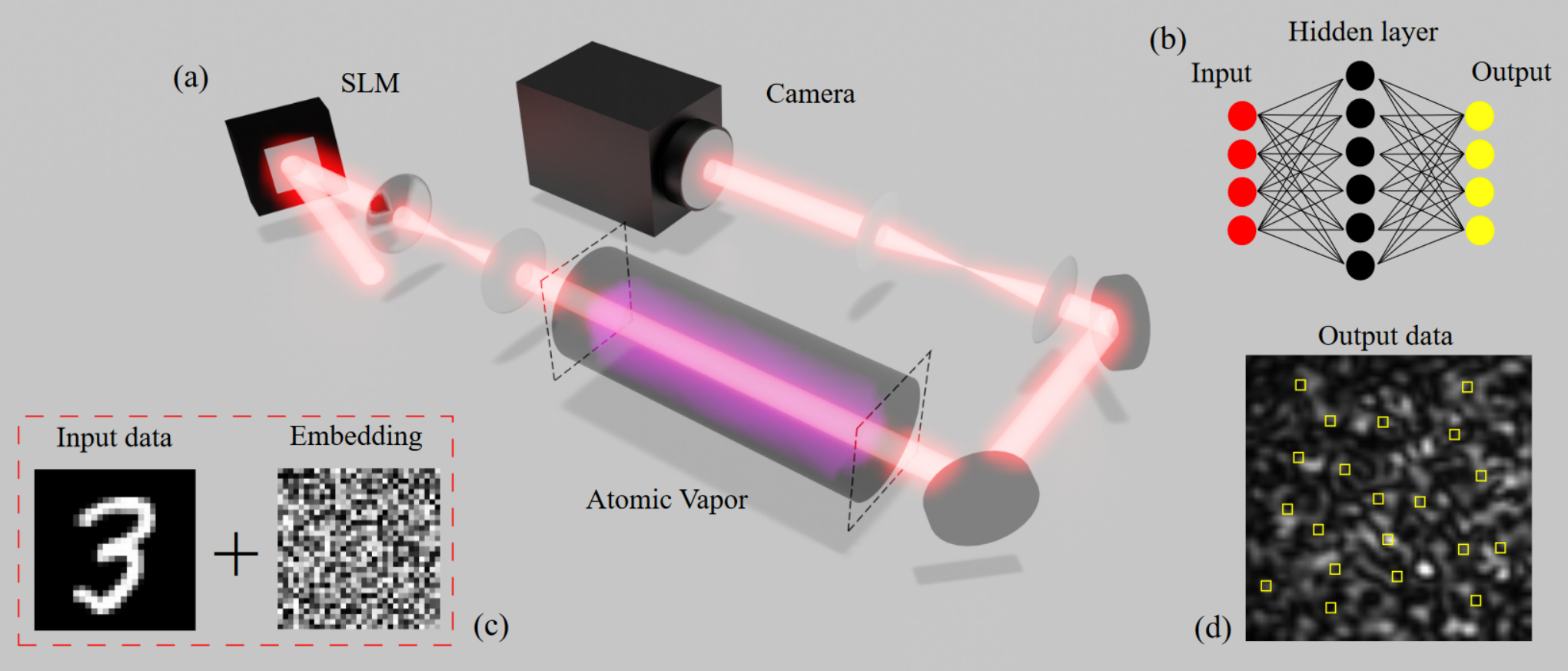}
\caption{Scheme and description of the optical system and its use as an Extreme Learning Machine Device. (a) The optical setup: The input data is encoded onto the beam with a SLM imaged at the entrance of the vapor cell. The data then propagates through the tunable nonlinear atomic vapor cell. The output image is recorded on a camera, conjugating the output plane of the vapor cell. (b) Extreme Learning Machine model. (c) Typical input data of MNIST dataset and a random embedding matrix added to the data. (d) Typical image recorded after nonlinear propagation of a MNIST data. Yellow squares are randomly distributed areas used as training parameters with associated weights ($\alpha$).}
\label{ML_scheme}
\end{figure*}

\noindent
learning tasks through numerical and experimental results highlighting the impact of such nonlinear medium on the learning efficiency. We then compare our system to an usual convolutional neural network (CNN), before concluding and providing a route for further improvement.
\section{Optical ELM architecture}

\noindent
A sketch of our hardware device is shown in Fig.\ref{ML_scheme}(a): a SLM is used to encode the data onto the incoming laser beam which then propagates through a hot atomic vapor (natural mixture of Rubidium) undergoing a nonlinear spatial transformation. The modified data is then recorded by a camera where the learning is numerically processed. The equivalence with a regular ELM scheme is illustrated in Fig.\ref{ML_scheme}(b): the SLM encodes $N$ samples ${\bf X} = [{\bf X}_1,...,{\bf X}_N]$ (inputs) in the optical system and the atomic vapor plays the role of the hidden layer (reservoir) that performs the data transformation ${\bf H}({\bf X})$. Finally, the camera records the resulting field and a linear combination with a set of $M$ weights $\boldsymbol{\alpha}$ is performed to get the outputs ${\bf Y = H} \boldsymbol{\alpha}$. The training of the system is then done by optimizing these weights by digitally solving a ridge regression problem \cite{HUANG12}.\\
The detail of our hidden-layer matrix ${\bf H}$ can be explained as follows: it projects input data into a feature space necessary to properly learn. In our system this projection is done completely optically and can be written as:
\begin{eqnarray}
    {\bf H} = {\bf D} [{\bf A}(p({\bf X},{\bf W}))]
    \label{feature mapping}
\end{eqnarray}

\noindent where ${\bf D}$ is the detection function of the camera, {\bf A} is a complex function modeling the spatial evolution of the beam along its propagation, $p$ is the encoding function of the SLM and ${\bf W}$ is an embedding matrix added to the encoding and creating biases on the input that leads to an enhancement of transformations during the propagation of the data.\\
The embedding ${\bf W}$ in our optical system is implemented by adding to each samples the same random matrix as presented in Fig.\ref{ML_scheme}(c). The encoding as mentioned above uses a SLM to imprint the information onto the phase profile of the laser beam: $p({\bf x})=\exp (i{\bf x})$. Considering a gaussian beam of waist $w$ propagating along $z$, the laser field is then:

\begin{eqnarray}
\psi({\bf r},z)=\sqrt{I}\exp\left(-\frac{r^2}{w^2}\right)\exp{i ({\bf X(r)} + {\bf W(r)})}
\end{eqnarray}

\noindent where $I$ is the beam intensity at the center, ${\bf r}=(x,y)$ ($r=\sqrt(x^2+y^2)$) is the coordinate in the transverse plan of propagation of the beam and $w$ is the waist of the laser beam. It is important to note that the use of complex values transformed in a continuous space largely enhances the performance of the model compared to a real valued binary solution.
The detection function ${\bf D}$ can also increase the complexity of the hidden layer by 

\begin{figure*}[!t]
\centering
\includegraphics[width=1.0\linewidth]{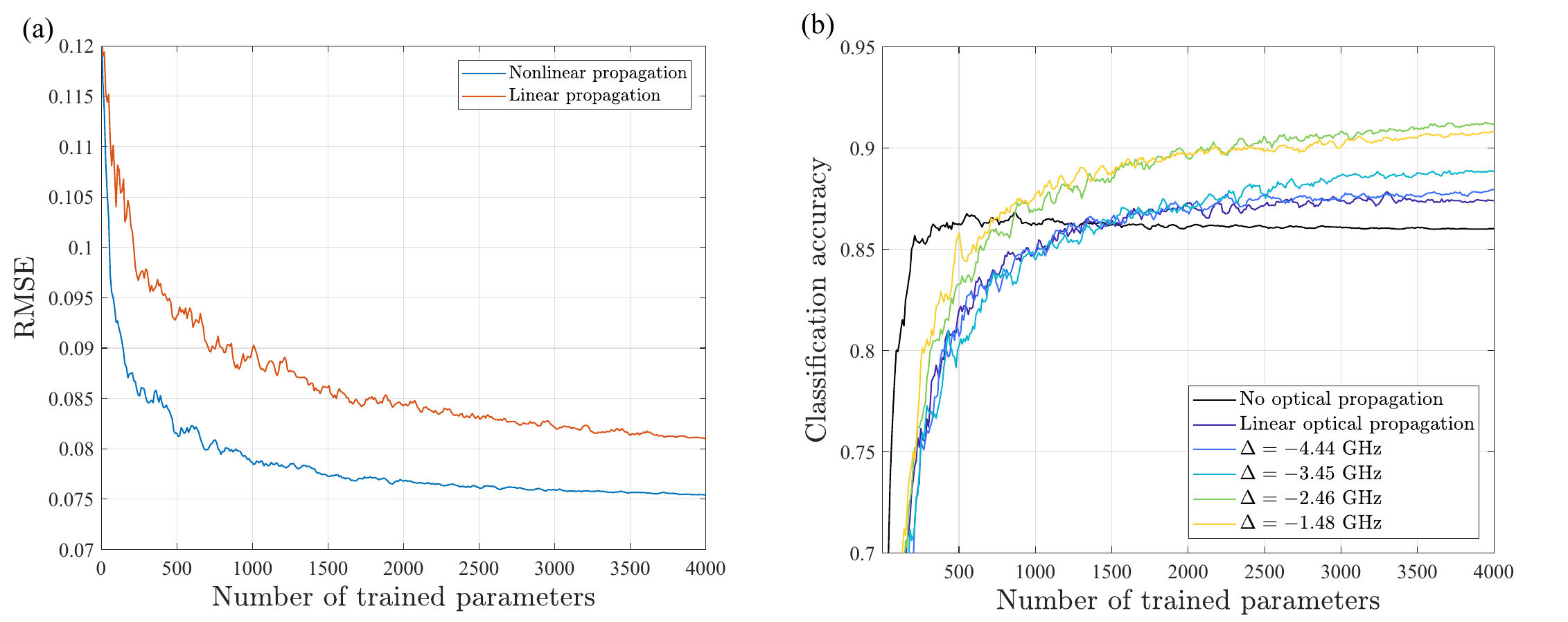}
\caption{Classification error and accuracy as a function of the number of training parameters. (a) Classification error (RMSE) for a numerical simulation with Abalone dataset ($4\,000$ data) comparing a linear (red) and a nonlinear (blue) propagation. (b) Classification accuracy for MNIST datasets : numerical results 
for a linear regression on the generated input mask, corresponding to no optical propagation (black line) and experimental results in presence of optical propagation: linear optical propagation (violet) and optical propagation with different laser frequency detunings (i.e. nonlinearities). Reducing $|\Delta|$ increases the nonlinearity. Curves have been smoothed for better readability.}
\label{training results}
\end{figure*}

\noindent
using the saturation properties of the camera, with e.g. a linear response up to a threshold intensity value given by $I_s$. The detection function can be approximated by ${\bf D({\bf I})} = min({\bf I}, I_s)$ with ${\bf I}$ the intensity matrix incident on the camera.
The hidden-layer matrix of our optical system can thus be rewritten as:

\begin{eqnarray}
    {\bf H} = D({\bf I}) [{\bf A}[\psi({\bf r},z)]]
    \label{feature mapping optical}
\end{eqnarray}

\noindent Finally, the propagation of the data through the nonlinear vapor is defined by the transformation ${\bf A}$ and is described by the nonlinear Schr\"odinger (NLS) equation. In the paraxial approximation for the optical field propagating along $z$ through the nonlinear medium, the slowly varying complex amplitude $\psi({\bf r},z)$ of the laser field is described by:

\begin{eqnarray}
    i \frac{\partial \psi ({\bf r},z)}{\partial z}=\left(-\frac{1}{2k_0}\nabla_{\bf r}^{2} - k_0 n_2 |\psi({\bf r},z)|^2 \right ) \psi({\bf r},z)
    \label{NLSE}
\end{eqnarray}

\noindent where $k_0$ corresponds to the wave vector of the laser beam. The last term of the equation shows that the nonlinear propagation of the light inside the vapor depends on the intensity of our beam $I = |\psi({\bf r},z)|^2$. Here $n_2(\rho, \Delta)$ is the first order nonlinear refractive index, $\rho$ is the atomic density of our vapor (directly linked to its temperature) and $\Delta$ is the frequency detuning between the laser and the atomic resonance (with respect to the $F=2 \rightarrow F'$ transition of $^{87}$Rb). It is thus possible to control 
the nonlinear transformation seen by the data by modifying the beam intensity, its frequency or the temperature of the vapor cell. In the context of machine learning, these control parameters can be defined as hyperparameters.\\
Once the optical transformation performed and recorded by the camera, the learning step can be processed digitally. As presented on Fig.\ref{ML_scheme}(d), $M$ channels (yellow squares) are randomly selected on this output matrix, corresponding to training parameters and a weight $\alpha$ is assigned to each of them. These channels are assimilated to training parameters. In order to reduce technical noise, each channel is an average over a square of $8\times8$ pixels. The digital learning is a simple ridge regression task, in our case processed using Scikit-learn \cite{pedregosa11} on Python.

\section{Results}

We first illustrate the impact of a nonlinear propagation by a numerical investigation. We therefore use samples from the Abalone dataset, which we imprint onto the phase profile of a gaussian beam whose propagation is simulated numerically.
We compare two different situations: one with a pureley linear propagation and one using the nonlinear transformation as described by the NLS equation. The learning task is then performed identically for the linear and nonlinear propagation. We present the corresponding regression accuracy by plotting the root mean square error (RMSE) in Fig.\ref{training results}(a) on a test set as a function of the number of training parameters obtained in this two cases.
This numerical simulation illustrates the positive impact that a nonlinear propagation can have on the learning efficiency of such an ELM model. We find that a weak nonlinearity is already sufficient to observe an enhancement of the inference accuracy from $91.8\%$ to $92.5\%$. It is also interesting to notice a particular strength of ELM models, with the regression converging toward a high accuracy with relatively few training parameters only ($\approx 1\,000$). We stress that the efficiency of ELM models to learn from small number of parameters leads to a potentially important gain in terms of time and energy of training. 

We have however realized that the computation time needed to simulate our optical system is already extremely large even for a relatively small dataset like Abalone. We therefore continued our investigation on larger datasets directly on our optical platform. Indeed as our free-space optical system is well suited for image data processing and ELM is a good model for classification, the experimental study has been performed for a classification task using the larger MNIST dataset composed of $28\times28$ pixels images of handwritten digits. 
The results of the corresponding classification accuracy is shown in Fig.\ref{training results}(b). First, we show as a reference the case of purely digital learning, in the absence of any optical encoding and propagation. 
The corresponding black line in Fig.\ref{training results}(b) illustrates the fast learning (with less then 1000 trained parameters), which however comes along a limited accuracy of $86.0\%$. Using an optical free space propagation already allows for an increased accuracy of $87.4\%$, underlying the impact of the encoding and recording optical devices (i.e. encoding and detection function $p$ and $D$ in eq.\ref{feature mapping}).
We note that as discussed earlier, a random embedding matrix has been added to the sample to enhance interactions during the free space beam propagation. We have checked that using different random matrices leads to significant fluctuations on the classification error, implying that the embedding matrix needs to be optimized for specific tasks. We also noticed that the number of pixels per channel has an impact on the accuracy, with a smaller number of pixels per channels improving the learning efficiency.
Concerning the recording, it is important to take into account the saturation of the camera described in Eq.\ref{feature mapping optical}. Preliminary experimental classification tests with different values of saturation parameter confirmed that a lower saturation intensity $I_s$ of the camera enhances the accuracy of the model. After optical propagation and recording on a CCD, the learning protocols is the same as for the purely digital process and involves associating a weight to each training parameter, consisting to a group of pixels (channel) of the CCD. We note that both numerical and experimental study presents a similar increase of accuracy with the number of training parameters. Even though the fully digital ELM model stays at lower accuracy, it converges with fewer trained parameters. We associate this difference to a spreading of the information of the input matrix during free-space propagation requiring more output channels to access all the information spread over more output channels.

We now turn to the characterization of the nonlinear beam propagation through the atomic vapor, the main novelty of our optical platform. As explained earlier, the nonlinear strength of the atomic vapor depends on three experimental hyperparameters, laser intensity, laser frequency (detuning) and atomic vapor density. In the results presented in this paper, we keep the laser intensity fixed at $I\approx1.3\text{ W.cm}^{-2}$. The linear beam propagation regime is easily reached by using the atomic vapor cell at room temperature, where the low atomic density results in no noticeable impact of the atoms on the index of refraction. The nonlinear propagation regime is obtained when heating the atomic vapor cell. We then control the nonlinearity more conveniently via the laser detuning $\Delta$, as tuning the laser frequency closer to the atomic transition frequency increases the nonlinear strength of the medium. We note that in all studies presented in this paper, we have used a negative detuning, corresponding to a defocusing nonlinearity.
In Fig.\ref{training results}(b) we observe an enhancement of the learning capacities when increasing the nonlinearity (i.e. reducing detuning $|\Delta|$): the classification accuracy with $4\,000$ training parameters for increasing nonlinearity is $88.0\%$, $88.8\%$, $91.2\%$ and $90.7\%$. It is interesting to notice a decrease of efficiency between the strongest nonlinearity $\Delta = -1.48$ GHz and the one before $\Delta = -2.46$ GHz. It is well known that optical propagation in a strong nonlinear medium can exhibit beam cleaning characteristics \cite{krupa17}, which can lead to smoothen out some of the information imprinted onto the beam resulting in a decrease of learning efficiency. This effect, connected to condensation of photons under nonlinear propagation, is more prominent in optical fibers \cite{Psaltis21} and appears after longer nonlinear propagation for free space devices as used in our platform \cite{Santic2018}. Additional effects have been observed in hot atomic vapors such as light absorption, non-locality and saturation of the nonlinearity \cite{Azam21} and remain to be explored in the context of machine learning efficiency. A full exploration to determine optimal experimental hyperparameters giving access to a nonlinear regime that maximizes the learning efficiency thus remains to be performed.
\\

\begin{figure*}[!t]
\centering
\includegraphics[width=1.0\linewidth]{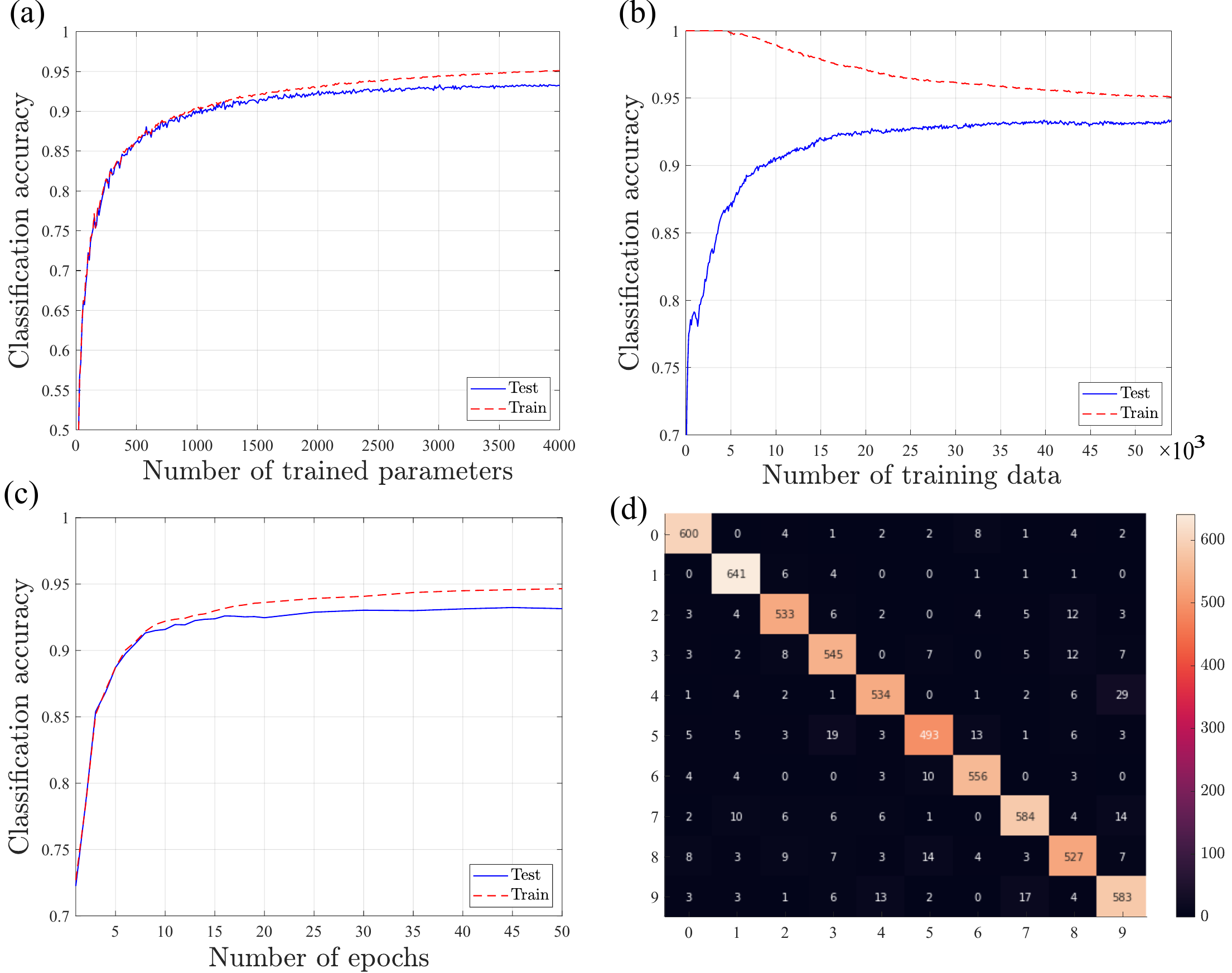}
\caption{Experimental training and inference classification accuracy for different hyperparameters on $60\,000$ samples of the MNIST dataset ($54\,000$ for the training and $6\,000$ for the testing). (a) Accuracy as a function of the number of trained parameters with the full dataset. (b) Accuracy as a function of the number of training data of the MNIST dataset with $4\,000$ trained parameters. The number of testing data is fixed at $4\,000$ samples. (c) Accuracy as a function of epochs with the full dataset and $4\,000$ trained parameters. (d) Confusion matrix obtained in the best case: full dataset, $4\,000$ trained parameters and $50$ epochs.}
\label{60k}
\end{figure*}

The study presented on Fig.\ref{60k} has been performed on the full MNIST dataset with nonlinear propagation ($\Delta = -2.46$ GHz) to better understand the training and testing behavior of the model for larger datasets. Similarly to the previous discussion, Fig.\ref{60k}(a) shows the evolution when increasing the number of trained parameters. Once again, it highlights that good accuracies can be achieved with very low number of training parameters, an intrinsic advantage of ELM models. We can also observe on Fig.\ref{60k}(b) that the accuracy of our optically accelerated ELM converges towards its final value with only a third of the dataset. 

The last study presented on Fig.\ref{60k}(c) corresponds to the evolution of the accuracy as a function of the number of epochs (i.e. times the digital training has been performed on each samples, after their optical propagation). This curve shows that a tens of epochs are enough to reach a final value. Finally, Fig.\ref{60k}(d) presents the confusion matrix of the best case using the full dataset, $4\,000$ training parameters and 50 epochs, leading to a classification accuracy of $93.25\%$ for the testing and $95.10\%$ for the training. On top of presenting the best accuracy reached so far with our ELM photonic platform, Fig.\ref{60k} highlights the low requirements needed by the model to perform a good testing accuracy. 

\section{Comparison with CNN and future prospects}

With this first benchmarking for machine learning performed with our hot atomic vapor as an extreme learning machine, we now compare the potential of this hardware to other standard machine learning systems. We therefore present in Table \ref{Table_Lenet5} a comparison between a simple convolutional neural network (Lenet-5) \cite{lecun15} and our system. For this CNN model, we have used $3$ convolutional $+$ $2$ average pooling layers, well suited to perform MNIST classification. Here, we have expanded the samples and used an adapted version of Lenet-5 to study the impact of sample size on CNN performances. By resizing MNIST 

\begin{table*}[!]
\centering
\begin{tabular}{ c|c|c|c }

Platform /  & Number of  & Time per sample  & Testing accuracy \\
Model (input size) &  required parameters & (ms)  &   \\
\hline
CPU / Lenet-5 ($28 \times 28$) & $60\,000$ & $1.6$  & $94.50\%$ \\ 
\hline
CPU / Lenet-5 ($64 \times 64$)  & $340\,000$ & $5.0$  & $94.40\%$ \\ 
\hline
CPU / Lenet-5 ($128 \times 128$)  & $1\,740\,000$ & $19.0$  & $89.20\%$ \\ 
\hline
CPU / Lenet-5 ($1000 \times 1000$)  & $118\,100\,000$ & $1100.0$  &  unknown \\ 
\hline
NLO / ELM ($28 \times 28$) & $4\,000$  & $20$ & $91.20\%$ \\
\hline
NLO / ELM ($1000 \times 1000$) & $\sim 1\,000\,000$  & $20$ & unknown \\ 

\end{tabular}

\caption{MNIST classification performances (number of trained parameter, speed, accuracy) using Lenet-5 (CNN) with different sample sizes and our nonlinear optics (NLO) device as an ELM. "Input size" corresponds to the image resolution (in pixels) sent into the model. The accuracies have been obtained after trainings over $10\,000$ samples. The CPU used to perform this comparison is an Intel core i5-$4690$ $3.5$GHz.}
\label{Table_Lenet5}
\end{table*}

\noindent
samples up to $128 \times 128$, both the number of training parameters required by the digital model as well as the time it takes to process a sample through the $5$ layers increase drastically. The accuracy presents also a tendency to decrease when increasing the sample size. Even though the CPU we have used in this example is not the most efficient, this test highlights a common problem of CNN where an increase of sample size has critical consequences in terms of time and energy of the required computation.

In the first 2 columns of the table we show how the number of required training parameters and the required computation time  scale with the samples size in the case of a learning using an adapted version of Lenet-5 (CNN) and for our ELM model. Both the number of required training parameters and the computation time scale as the size size $N \times N$. We stress that the reduced number of training parameters for our approach is an inherent advantage of all ELM models.
The third column shows the required computation time. Again, for the  Lenet-5 (CNN) model, one can see that the computation time scales again linearly with the size size $N \times N$. However, when using our nonlinear optical approach to implement the ELM model, we can see that the computation is independent of the system size! This present a mayor advantage for our nonlinear optical acceleration on top of the use of the ELM model. We note that the ELM model requires less training parameters allowing to scale the size of the system to $1000 \times 1000$, whereas for the CNN based approach this larger number of required training parameters put very hard constraints on the computational effort.
The last column of the table indicates the accuracy of the different situations, showing the reduced accuracy for increasing system size in our implementation of the Lenet-5 (CNN) model. We note that due to restricted computation time, we were not able to perform the full training on the $1000 \times 1000$ sample size. Similarly, for our nonlinear optical approach, we did not yet implement a  $1000 \times 1000$ sample (which required further development in our hardware). We note that in the Lenet-5 model the number of required training parameters scales linearly with the input size whereas in our platform the input size is dependent on the SLM resolution while the number of training parameters depends on the CCD resolution. We stress that, in our system, the computation time is the same for large and small samples, an important asset for any optical solver. 

In summary,  this table thus shows that a simple CNN is more efficient than our present system for small samples (such as the MNIST dataset used in this paper). We note that the limitation in computational time of our optical platform is stemming from the electrical to optical (SLM) and optical to electrical (camera) transformations. However, the main advantage of using a free-space optical platform is the possibility of processing very large samples without supplementary cost in computation time. Indeed, these devices are able to process at the same speed of $50$Hz samples of size up to $10^6$ pixels (with commercially available SLMs already reaching sampling times of $100$ Hz / $10^7$ pixels). 
In order to explore the potential of our nonlinear optics platform, let us assume that each pixel of the encoding device can be assimilated to an optical mode propagating along the optical path and is interacting with the other modes via the nonlinear atom-photon interaction occurring inside the hot atomic vapor. Such interaction between $2$ pixels being equivalent to a mathematical operation in a digital processing unit, we estimate that our system (with the actual encoding technology) is able to perform $10^{12}$ 8-bit operations per frame at a rate of $50$Hz, leading to potentially $5\cdot 10^{13}$ OPS (operations per second). With an energy consumption of our entire system (including ELM processed digitally and the heating of the vapor cell) of about $500$W, the efficiency of our platform can be estimated to be of the order of $10^{11}$ OPS/W (operations per second per watt). We note that the cost (in time and energy) of the learning part of this protocol itself remains negligible. Indeed the inherent properties of ELM models needs only a simple pass through only one layer of training parameters, whose number is always kept low. Based on the results of our first experimental study with the MNIST dataset, we estimate that a number of training parameters (channels on the output image) equivalent to the input size ($L \times L$) should be sufficient to reach an accuracy of at least $90\%$. 
It is also worth noting that this time and energy advantage of our nonlinear optical platform does not come at the expense of increased fabrication cost. If the expected performances of our platform become competitive to present day supercomputers (used for machine learning tasks) our platform can be realized with a hugely reduced fabrication cost and hardware size. Indeed, while the actual best supercomputer Frontier \cite{schneider22} costs $600$ millions US dollars to build and consists of $74$ $19$-inch rack cabinets (over $680$ m$^2$), we estimate that our solution could roughly cost a few $100$ thousands of US dollars and the entire system being contained into one $19$-inch rack cabinet. This small size also provides an important advantage for tasks requiring on-site or off-line work.\\
We stress that the results presented in this paper consist in a very first proof of principle and our device is far from being optimized for machine learning purposes. 
The embedding matrix for instance plays a key role on how the data is projected into the feature space and it will be interesting to explore how different matrices or encoding methods would impact the learning efficiency. 
We had e.g. implemented randomly distributed embedding matrices to maximize the connections into the sample along its optical propagation. Even if we did not observe any significant difference between adding or multiplying the sample with this embedding matrix, we noticed an important difference of accuracy from one random matrix to an other: a variation of almost $10 \%$ has been observed just by changing the seed of the random matrix. This difference when using various random distributions of this embedding matrix suggests a potential for optimization which is likely to depend on the task and the dataset the system is used on.
Similarly as \cite{oguz24, Wu22}, a learning step has to be implemented on the optical encoding to optimize the impact of the embedding matrix.
Even though the above technique is already used by other optical platforms, our nonlinear atomic vapor gives the possibility to develop and enhance this technique. For the study presented in this paper, we encoded the information onto the phase profile of our laser beam alone, keeping its intensity, which is one of the nonlinear hyperparameters, as a constant. It is however possible to use a SLM to simultaneously control both intensity and phase profile of a beam \cite{bolduc13} via the use of a Blazed grating, allowing to simultaneously encode the sample onto the phase beam with the embedding matrix and locally tune the nonlinear strength of the atomic vapour via the use of the intensity profile. Therefore, the supplementary step needed to optimize the embedding matrix (phase profile) will also be used to determine the appropriate intensity profile enhancing the nonlinear transformation for a specific task or dataset.\\

\section{Conclusion}

Optical platforms have been used in the past to demonstrate various advantages they offer to perform for complex calculations tasks such as machine learning. At present, many new systems are being developed in order to determine which ones will reach competitive performances against fully digital solutions. With this objective in mind we have presented in this work the use of an optical hardware based on nonlinear light propagation through a hot atomic vapor and its integration into a Extreme Learning Machine model. Combining the large scalability of optical free-space propagation with the strong and tunable nonlinearity of our optical hardware with the specific design of ELMs optimized for fast learning on large and heavy datasets could lead to important savings in terms of time and energy of the computation. In this paper, we have presented first evidence of learning enhancement via the use of a hot atomic vapor as a nonlinear medium combined with a extreme learning machine model in the context of image classification task. This study, performed on the MNIST dataset as a benchmark tool, shows that reasonable accuracies can be achieved quite easily. More importantly, it highlights the potential of optical platforms both in term of scalability and efficiency regarding the time and energy cost of large samples.
We note that other promising optical platforms using e.g. multimode fibers propagation to process the nonlinear transformation \cite{Psaltis21, ancora22} are limited on this aspect due to the low number of mode they can inject into the fibers.
Various applications using large data (large language model and image recognition for example) being currently developed, there is a necessity to create new machine learning platforms in order to efficiently reduce the cost of such applications. When fully developed, such platforms could become competitive to the actual best supercomputers in terms of speed and energy cost of calculation along with lower fabrication cost and smaller hardware size. Finally, future implementations are envisioned such as the learning of the phase embedding matrix and a local tuning of the nonlinear medium via an adapted intensity matrix to drastically enhance the accuracy of the platform for various kind of tasks.

\section{Acknowledgements}
We thank M.Morisse for fruitful discussions on machine learning and C.Conti and D.Pierangeli for their discussions on optical extreme learning machine. We also thank C.Jurczak for his support throughout this project and CDL (Creative Destruction Lab) for their advise and mentorship.
We also acknowledge funding from CNRS Innovation via the prematuration project SQVAC and Provence-Alpes-Côte d'Azur région via a 'Jeune docteur innovant' project.

\section{Supplementary material}

\subsection{Experimental setup}

The platform uses a continuous laser beam at $780$nm with a gaussian profile of waist $w=10$mm and a power fixed at $1.3$W. Once linearly polarized, this beam illuminates a Spatial Light Modulator (SLM) from Hamamatsu (model X15223), encoding 8-bit values on $1024 \times 1242$ pixels with a pitch of $12.5$ $\mu$m and at a frame rate of $50$ Hz. The SLM encodes onto the beam the desired matrix by imposing local phase delay of the beam from $0$ to $2\pi$ with $255$ values. The signal displayed by the SLM is the sum of the sample and the embedding matrix. The plan of the SLM is then imaged via an afocal system at the entrance of the $7$cm vapor cell. The vapor, a natural mixture of Rubidium ($72\%$ of 85Rb and $28\%$ of 87Rb), is heated at $120^\text{o}$C to increase atomic density (to $\sim10^{19}at/m^3$), enhancing photon-atom interactions. The laser frequency is detuned by several GHz toward the red from the atomic resonance in order to optimize the effect of the Kerr-like nonlinearity. In such regime the vapor behaves as a defocusing medium, this effect being enhanced by the beam intensity itself, we talk about the self-defocusing phenomenon. Once the beam has propagated through the vapor, the output window of the cell is imaged onto a CCD camera. In this first study, to avoid noise, the image displayed by the SLM is magnified by a factor $10$ and the decoding channels used to learn are an average over $10 \times 10$ camera pixels.\\
All the samples are processed one by one through the optical system with fixed hyperparameters (detuning, vapor temperature, beam power, camera saturation,..) and embedding matrix. Once recorded by the camera they are stored on the computer controlling the system. Randomly distributed channels are chosen to perform the learning on the training dataset by determining the weight of all these channels. When the training is finished, the channels with the adequate weights are used to process the classification on the testing dataset.\\

\subsection{Datasets}

The Abalone dataset \href{https://www.kaggle.com/datasets/rodolfomendes/abalone-dataset}{( https://www.kaggle.com/data sets/rodolfomendes/abalone-dataset )} is a commonly used dataset to benchmark machine learning hardwares / models on regression task and is well suited to start working on new systems due to its low number of samples ($8\,177$). This dataset concerns the identification of sea snails, each of them having $8$ attributes (age, length, diameter, height, rings,...).\\
The MNIST dataset \href{https://www.kaggle.com/datasets/hojjatk/mnist-dataset}{( https://www.kaggle.com/datasets /hojjatk/mnist-dataset )} concerns a classification task on a multi-class problem: handwritten digits. The dataset includes $10$ classes from $0$ to $9$ over $60\,000$ samples for the training and $10\,000$ for the testing. Each of these samples are matrices of $28 \times 28$ pixels with filled $8$-bit values. This widely used dataset is a necessary step to benchmark a machine learning platform on classification tasks. The 2D format of the samples makes it a perfect first playground for systems aiming to develop image based tasks.

\end{document}